\documentclass[reprint,showpacs,amsfonts,amsmath,amssymb,prb]{revtex4-1}
\usepackage{graphicx}
\usepackage{dcolumn}
\usepackage{bm}
\usepackage{color}

\begin{document}

\title{Quasiparticle scattering interference in iron pnictides: A probe of the origin of nematicity}
\date{\today}

\begin{abstract}
In this paper, we investigate the quasiparticle scattering
interference(QPI) in the nematic phase of iron pnictides, based on
the magnetic and orbital scenarios of nematicity, respectively. In
the spin density wave(SDW) state, the QPI pattern exhibits a dimer
structure in the energy region of the SDW gap, with its
orientation along the ferromagnetic direction of the SDW order.
When the energy is increased to be near the Fermi level, it
exhibits two sets of dimers along the same direction. The dimer
structure of the QPI patterns persists in the magnetically driven nematic phase,
although the two dimers tend to merge together with energies
closing to the Fermi level. While in the orbital scenario, the QPI
patterns exhibit a dimer structure in a wide energy region. It
undergoes a $\pi/2$ rotation with the increasing of energy, which
is associated with the inequivalent energies of the two Dirac
nodes induced by the orbital order. These distinct features may be
used to probe or distinguish two kinds of scenarios of the nematicity.
\end{abstract}

\pacs{74.70.Xa, 72.10.Fk, 75.30.Fv, 75.25.Dk}
\author{Hai-Yang Zhang and Jian-Xin Li}
\email[]{jxli@nju.edu.cn}
\affiliation{National Laboratory of Solid State Microstructure and Department of
Physics, Nanjing University, Nanjing 210093, China\\
Collaborative Innovation Center of Advanced Microstructures, Nanjing University, Nanjing, China}
\maketitle
\date{\today}

\section{Introduction}

In recent years, the unconventional superconductivity found in
iron-based superconductors(IBSCs) has attracted much attention in
condensed matter community. Due to the proximity of
superconductivity and the collinear spin density wave(SDW) phase,
it is generally believed that there exists an intrinsic link
between the magnetic fluctuations and the superconductivity. Thus,
the understanding of the normal state magnetic fluctuations will
be helpful to identify the mechanism of superconductivity.
Experimentally, it was found that the SDW transition is either
preemptive by or coincident with a tetragonal-to-orthorhombic(TO)
structural transition at $T_{s}$\cite{BFCA,BKCA,NFA} which signals
the $C_{4}$ symmetry breaking above the SDW transition
temperarture $T_{N}$. Such a $C_{4}$ symmetry breaking phase is
called the nematic phase in literatures\cite{NMREV}. Further
studies show that the nematicity persists to $T^{*}$ above
$T_{s}$\cite{NATM,OODING,PNAS,FESE}. So far, there is clear
evidence that nematicity comes from the electronic rather than the
lattice degree of freedom\cite{SCIDZ,PRBDZ}. However, it is
difficult to identify the fingerprint of nematicity due to the
coupling of various electronic degrees of freedom. Experimentally,
it was found that the presence of nematicity is accompanied by
anisotropic magnetic correlations which indicates its magnetic
origination\cite{NATM,PCDAI}. Meanwhile, it was also found that
the characteristic temperature $T^{*}$ of nematicity coincides
with that of the orbital order\cite{OODING,FEPG,JPSJOM}, which indicates
that nematicity is orbitally driven. So, the origin of nematicity
is still debated, both the magnetic and orbital fluctuations are
proposed to be responsible for the emergence of
nematicity\cite{CKXU,CFANG,ONARI,FERNANDES}. Thus, the study of
the origin of nematicity is highly desired as it may shed light on
the mechanism of unconventional superconductivity in IBSCs.

One of the effective ways to distinguish between different
scenarios of nematicity is to compare the theoretical results of
each scenario to the experimental observations. In this regard, we
notice that the electronic structure can be mapped effectively
through the analysis of the quasiparticle scattering interference
(QPI) patterns\cite{Hoffman,WANG,Lee,Lee1} and the spatial modulation of
the electronic states resulting from QPI can be probed directly by
the STM experiments. In fact, the recent STM
experiments\cite{QPIN,QPIP} for NaFeAs have been performed to
investigate the electronic structure in the nematic phase in
IBSCs. The STM experiments\cite{QPIN,QPIP} reveal that the QPI
patterns exhibit highly anisotropic dimer structure deep in the
SDW phase. It was further found that the anisotropic features
persist to high temperatures above $T_{s}$, but the anisotropy
weakens gradually with the increase of temperature. Therefore, the
QPI may offer a playground for the test of the origin of
nematicity.

In this paper, we study theoretically the quasiparticle
interference induced by impurity in the nematic phase based on the
magnetic and orbital scenarios of nematicity, respectively. Deep
in the collinear $(\pi,0)$ SDW state, the low energy QPI patterns
exhibit a dimer structure with its orientation along the
ferromagnetic direction of the SDW state. This is due to the fully
opening of the SDW gap along the $k_x$ direction. When the energy
is increased to be slightly below the Fermi level, it is composed
of two sets of dimers along the same direction. In this case, the
QPI reflects the topology of the distorted Fermi surface and the
two sets of dimers come from the inter-hole-pocket and
intra-hole-pocket scatterings. It is further shown that the above
features of the QPI patterns remain in the magnetically driven
nematic phase, which is modelled by the fluctuating short-range
antiferromagnetic order. In the orbital scenario of nematicity, it
is found that the scatterings of quasiparticles between Dirac cones
dominate the QPI process. Due to the
inequivalent energy positions of the two Dirac cones resulting
from the $C_{4}$ symmetry breaking induced by the orbital order,
%the QPI patterns undergoes a transition from the horizontal to vertical structure with increasing energy.
the dimer structure in the QPI patterns undergoes a $\pi/2$
rotation with increasing energy. The transition is irrespective of
the momentum dependence of the orbital order, so long as the
energy splitting between the $d_{xz}$ and $d_{yz}$ bands is fixed
near the Dirac cones. We propose that these results can be used to
distinguish the origin of the nematicity in various iron-based
compounds.

As a preliminary comparison to available experiments, we note that
the obtained anisotropic features of the scattering patterns and the
orientation of the dimer in the magnetic scenario
are qualitatively consistent with the experimental observations on NaFeAs\cite{QPIN,QPIP}.

%This paper is organized as follows. The model and formulas are described in Sec.II. We calculate the experimental consequences within the magnetic and orbital scenarios of nematicity,
%and compare the theoretical results with the experimental observations in Sec.III. In the last section, we present brief summary and discusses.

\section{model and formulas}
%To carry out the calculation, we adopt the five-orbital tight
%binding Hamiltonian of Ref\cite{GRASER} which reproduces the LDA
%energy bands. The tight binding Hamiltonian reads
%$H_{0}=\sum_{k,a,b,\sigma}
%\epsilon_{ab}(k)C^{+}_{a\sigma}(k)C_{b\sigma}(k)$. Where $a,b$ and
%$\sigma$ are the orbital and spin index, respectively. The tight
%binding hopping parameters for $\epsilon_{ab}(k)$ are given in
%Ref\cite{GRASER} and the energy unit eV will be used throughout
%this paper. In the present case, we focus on the QPI phenomenon
%induced by impurity potential scattering. The impurity Hamiltonian
%can be written as $H_{imp}=V_{ab}
%C^{+}_{i_{0}a\sigma}C_{i_{0}b\sigma}$. For simplicity, we adopt a
%$\delta$-function type scattering potential which resides on the
%given lattice site $i_{0}$. The scattering matrix
%$V_{ab}(k,k+q)=V\delta_{ab}$ is orbital diagonal and momentum
%independence.

The Hamiltonian we use to carry out the calculations can be divided
into three parts: the tight binding part, the impurity part and
that modelling the nematicity.

We adopt the five-orbital tight binding Hamiltonian of
Ref\cite{GRASER} which reproduces the LDA energy bands. The tight
binding Hamiltonian reads $H_{0}=\sum_{k,a,b,\sigma}
\epsilon_{ab}(k)C^{+}_{a\sigma}(k)C_{b\sigma}(k)$. Where $a,b$ and
$\sigma$ are the orbital and spin indices, respectively. The tight
binding hopping parameters for $\epsilon_{ab}(k)$ are given in
Ref\cite{GRASER} and the energy unit eV will be used throughout
the paper. In this paper, we focus on the QPI phenomenon induced
by impurity potential scattering. The impurity Hamiltonian can be
written as $H_{imp}=V_{ab} C^{+}_{i_{0}a\sigma}C_{i_{0}b\sigma}$.
For simplicity, we adopt a $\delta$-function type scattering
potential which resides on the given lattice site $i_{0}$. The
scattering matrix $V_{ab}(k,k+q)=V\delta_{ab}$ is orbital diagonal
and momentum independent. In this paper, the
average electron occupation number is fixed to be $6.0$ per unit cell which corresponds
to the undoped parent compound.

%We use the onsite Coulomb interacting Hamiltonian to model the SDW
%order and the magnetic driven nematicity. Meanwhile, the orbital
%orders are used to model the orbital driven nematicity. Both cases
%will be detailed in the following.}}
\subsection{Modelling of the SDW state}
In order to model the SDW and the magnetically driven nematic
phase, we include the Coulomb interaction Hamiltonian. So, the
full Hamiltonian reads $H_{m}=H_{0}+H_{imp}+H_{int}$, and
$H_{int}$ is given by,
\begin{eqnarray}
H_{int}&=&U\sum_{i,a} n_{ia\uparrow}n_{ia\downarrow}+U^{'}\sum_{i,a<b}n_{ia}n_{ib} \nonumber\\
&+&J\sum_{i,a<b} C^{+}_{ia\sigma}C^{+}_{ib\sigma^{'}}C_{ia\sigma^{'}}C_{ib\sigma} \nonumber\\
&+&J^{'}\sum_{i,a\neq b} C^{+}_{ia\uparrow}C^{+}_{ia\downarrow}
C_{ib\downarrow}C_{ib\uparrow},
\end{eqnarray}
Where $a$ and $b$ are the orbital indices. $U,U',J,J'$ are the
coefficients of the intraorbital interaction, interorbital
interaction, Hund-coupling, and pair hopping terms, respectively.
$U=U^{'}+J+J^{'}$ and $J=J^{'}$ are assumed as required by the
spatial rotational symmetry. Without loss of generality,
$J=U/4$ is assumed. $H_{int}$ is treated at the mean field level,
and is decoupled into the orbital diagonal channel in the
following way\cite{BASCONES,FSLUO,ANDERSEN},
\begin{eqnarray}
H_{int}&\simeq& U\sum_{i,a,\sigma}\langle n_{ia\sigma}\rangle n_{ia\bar{\sigma}}+(U^{'}-\frac{J}{2})\sum_{i,a\neq b}\langle n_{ia}\rangle n_{ib} \nonumber \\
&-&2J\sum_{i,a\neq b}\langle\textbf{S}_{ia}^{z}\rangle\textbf{S}_{ib}^{z}.
\end{eqnarray}
Where $n_{ia}$ and $\textbf{S}_{ia}$ are the electron number and spin operators at site $i$ with orbital $a$, respectively.
%Only the orbital diagonal magnetic order is considered, thus the pair hopping term in $H_{int}$ does not %contribute to the mean-field Hamiltonian in the above mean-field treatment.

The values of $\langle n_{ia\sigma}\rangle$ and
the magnetic moment $\text{S}_{a}$ which is defined via $\text{S}_{a}=\frac{1}{N}\sum_{i}(-)^{i_{x}}\langle\textbf{S}_{ia}^{z}\rangle$ for the $(\pi,0)$ SDW state ($N$ the number of lattice sites, $i_x$ the $x$-coordinate of site $i$) are obtained through the
self-consistent calculations. In this paper, the intraorbital interaction $U=1.3$ is adopted, which is
slightly above the critical value $U_{c}=1.24$ for the appearance
of SDW. The results for the magnetic moments are
$\text{S}_{xz}=0.029$, $\text{S}_{yz}=0.071$, $\text{S}_{x^2-y^2}=0.024$, $\text{S}_{xy}=0.044$, and $\text{S}_{3z^2-r^2}=0.028$.
In Eq.(2), we have ignored the orbital off-diagonal magnetic terms. In fact, we have checked numerically the effects of these off-diagonal terms. We find that the nonzero
orbital off-diagonal SDW moments are $\text{S}_{x^2-y^2,3z^2-r^2}=\text{S}_{3z^2-r^2,x^2-y^2}=-0.017$, where $S_{a,b}=\frac{1}{2N}\sum_{i}(-)^{i_{x}}\langle C^{+}_{ia\alpha}\sigma^{z}_{\alpha\beta}C_{ib\beta}\rangle$. These SDW moments are smaller compared to their diagonal parts,
$\text{S}_{xz}=0.034$, $\text{S}_{yz}=0.077$, $\text{S}_{x^2-y^2}=0.029$, $\text{S}_{xy}=0.05$, and $\text{S}_{3z^2-r^2}=0.032$. Furthermore, we have checked that the off-diagonal SDW orders have no qualitative influences on the behaviors of the QPI patterns in the SDW and spin driven nematic state. Here, we focus mainly on the QPI within the orbital diagonal SDW approximation in the following discussions.

Within the $(\pi,0)$ ansatz of the SDW order, we introduce the
annihilation operator with ten components as
$\psi_{k\sigma}=(C_{ka\sigma},C_{k+Qa\sigma})^{T}$, where
$Q=(\pi,0)$ is the SDW wave vector. In this way, the Green's function reads
$G_{k\sigma}(\tau)=-\langle
T\psi_{k\sigma}(\tau)\psi^{+}_{k\sigma}(0)\rangle$ in the SDW state,
where $T$ is the time-ordering operator.
\begin{figure}
\centering\includegraphics[width=3.2in]{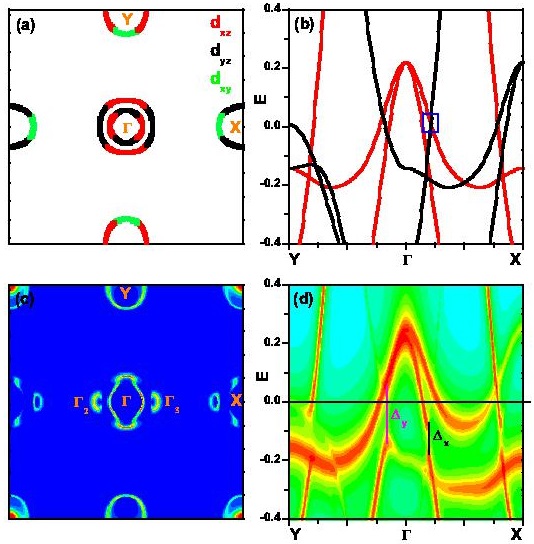}
\caption{(Color online) SDW distorted electronic structure. (a) The normal state
Fermi surface and its dominate orbital components without the SDW
order. (b) The red and black lines denote the original bands and
the folded bands by the SDW order with wave vector $(\pi,0)$,
respectively. The blue square marks the location of the Dirac node
in the momentum space. (c) The intensity map of the SDW distorted
Fermi surface. $\Gamma_{2}$ and $\Gamma_{3}$ are two Dirac
pockets. (d) The intensity map of the single-particle spectral
function along the high symmetry directions. The intensity is in
logarithmic scale in this panel.}
\label{f.1}
\end{figure}

\subsection{Modelling of the spin and orbital driven nematicity}
Experimentally, the nematicity in Fe-pnictides and
Fe-chalcogenides has been shown to develop at a temperature above the SDW transition\cite{NATM,OODING,PNAS,FESE}.
Therefore, the magnetically driven nematicity where the long-range SDW
order is absent but the C$_4$ symmetry is broken by magnetic
fluctuations has been proposed\cite{CKXU,CFANG,FERNANDES}. This
phase is modelled by the short-range antiferromagnetic
correlation, which is approximated by the Lee-Rice-Anderson
formula\cite{LEERICE}. In this way, the Green's function is
written as
$\tilde{G}^{-1}_{k}=i\omega_{n}-\hat{\epsilon}_{k}-\sum_{q}
P_{q}\frac{\hat{\Delta}^{2}}{i\omega_{n}-\hat{\epsilon}_{k+q+Q}}$,
where $\hat{\epsilon}_{k}$ and $\hat{\epsilon}_{k+q+Q}$ are the
matrix representation of the mean-field Hamiltonian $H_{m}$
without the SDW terms. $P_{q}=\frac{1}{\xi^{-2}+q^{2}}$ is a
Lorentzian which represents the $q$ modulated
magnetic correlation with $\xi$ the correlation length and $q$ the
momenta derivation from $(\pi,0)$. The orbital diagonal
$\hat{\Delta}$ is the order parameter matrix of the fluctuating
magnetic order. The elements of $\hat{\Delta}$ read as
$\hat{\Delta}_{aa}=(U-J)\text{S}_{a}+J\sum_{b}\text{S}_{b}$
and $\hat{\Delta}_{ab}=0$$(a\neq b)$. Where $\text{S}_{a}$ is
the magnetic moment of orbital $a$. Following the previous
study\cite{LEERICE}, the mean-field SDW order parameters
 obtained at $T=0$ are taken to be
$\text{S}_{a}$.

In the orbital scenario, the orbital orders are used to model the orbitally driven nematicity as used before\cite{ONARI,OOCHEN,OOLV}. Generally, the
Hamiltonian for the orbital orders can be written as
$H_{orb}=\sum_{ij,ab,\sigma}\lambda_{ij,ab}C^{+}_{ia\sigma}C_{jb\sigma}$
which breaks the $C_{4}$ symmetry, where $\lambda_{ij,ab}$ are the
order parameters. The value of
$\lambda_{ij,ab}$ is determined by the ARPES data. Thus, the full
Hamiltonian reads $H_{r}=H_{0}+H_{orb}+H_{imp}$ for the orbital
scenario of nematicity in the presence of the impurity scattering.
Both the momentum dependent and the momentum independent orbital
orders will be considered in the following.

With the Hamiltonians shown above, we can now construct the corresponding Green's functions.
For the orbital scenario of nematicity, the Green's function can be defined in the
orbital basis as $G_{ab}(\tau)=-\langle
T C_{ka}(\tau)C^{+}_{kb}(0)\rangle$ which can be obtained directly from $H_{r}$.

\subsection{Calculation method of the QPI}
The quasiparticle interference occurs between the ingoing and
scattered outgoing electrons by impurity. The resulting spatial
modulation of the electronic states can be visualized directly by
the STM experiments\cite{Hoffman,WANG}. After a Fourier
transformation, one can get its manifestation in the momentum
space, i.e., the density of states in the momentum space
$\rho_{q}(\omega)$. Thus, the features of the QPI patterns can be
qualitatively understood by the analysis of the joint density of
states of the initial and final states. Theoretically,
$\rho_{q}(\omega)$ is expressed as $\rho_{q}(\omega)=
-\frac{1}{\pi}{\rm Im} \sum_{k\sigma} {\rm
Tr}(G_{k\sigma}T_{k\sigma,k+q\sigma}G_{k+q\sigma})$, where the $T$
matrix reads
$T_{k\sigma,k+q\sigma}=V_{k,k+q}+\sum_{k^{'}}V_{k,k^{'}}G_{k^{'}\sigma}T_{k^{'}\sigma,k+q\sigma}$.
In each scenario, the corresponding Green's function $G_{k}$ is
used to calculate $\rho_{q}(\omega)$. A Born limited scattering
potential of $V=0.05$ is adopted. In this case, the QPI patterns
are not disturbed by the impurity resonance states, thus they are
directly related to the underlying bands structure.

\section{Numerical results}
\subsection{QPI patterns in the magnetic scenario of nematicity}
Let us start with the discussion of the distortion in the energy
bands due to the collinear $(\pi,0)$ SDW order with its moment
antiferromagnetically aligning along the $x$ direction but
ferromagnetically along the $y$ direction. In Fig. \ref{f.1}, we
show the normal state Fermi surface (FS) and the resulted FS after
the introduction of the SDW order. Clearly, the normal state hole
FS around $\Gamma=(0,0)$ develops into three distinct hole pockets
$\Gamma$, $\Gamma_{2}$ and $\Gamma_{3}$ in the presence of the SDW
order which induces the hybridization between the hole pockets
around $\Gamma$ and the electron pocket around $X=(\pi,0)$. As
revealed by previous study\cite{YRAN}, the two small hole pockets
$\Gamma_{2}$ and $\Gamma_{3}$ exhibit Dirac cone like dispersions.
These Dirac cone dispersions persist even when the off-diagonal SDW orders
are taken into account. The SDW distorted FSs calculated here are in qualitative agreement
with the ARPES observations\cite{APDING,APFENG,APSHEN,APZHANG}. At
the same time, the similar distortion of the Fermi pocket at $X$
occurs, while the Fermi pocket around $Y$ is affected less,
because the $(\pi,0)$ SDW order is considered here.
Quite similar SDW distorted Fermi surfaces have been obtained by the previous calculations
\cite{YRAN,FSLUO,FSGA} based on the five orbital model. Meanwhile, our results for the single-particle spectrum are fully consistent with the
previous study\cite{FSLUO}, as shown in Fig~\ref{f.1}(d).

%Because the $\gamma$ FS pocket is strongly   We also note there is a clear difference between the distorted FSs pockets around $X$ and $Y$.
%Due to the band folding effect, shadow bands develop around $X$.
%The coupling of the electron pocket around $Y=(0,\pi)$ and the hole band around $M=(\pi,\pi)$ weakens the spectral weight of $Y$ pocket. Shadow band of the $Y$ pocket and a rather small hole pocket emerge around $M$.
\begin{figure}
\centering\includegraphics[width=3.2in]{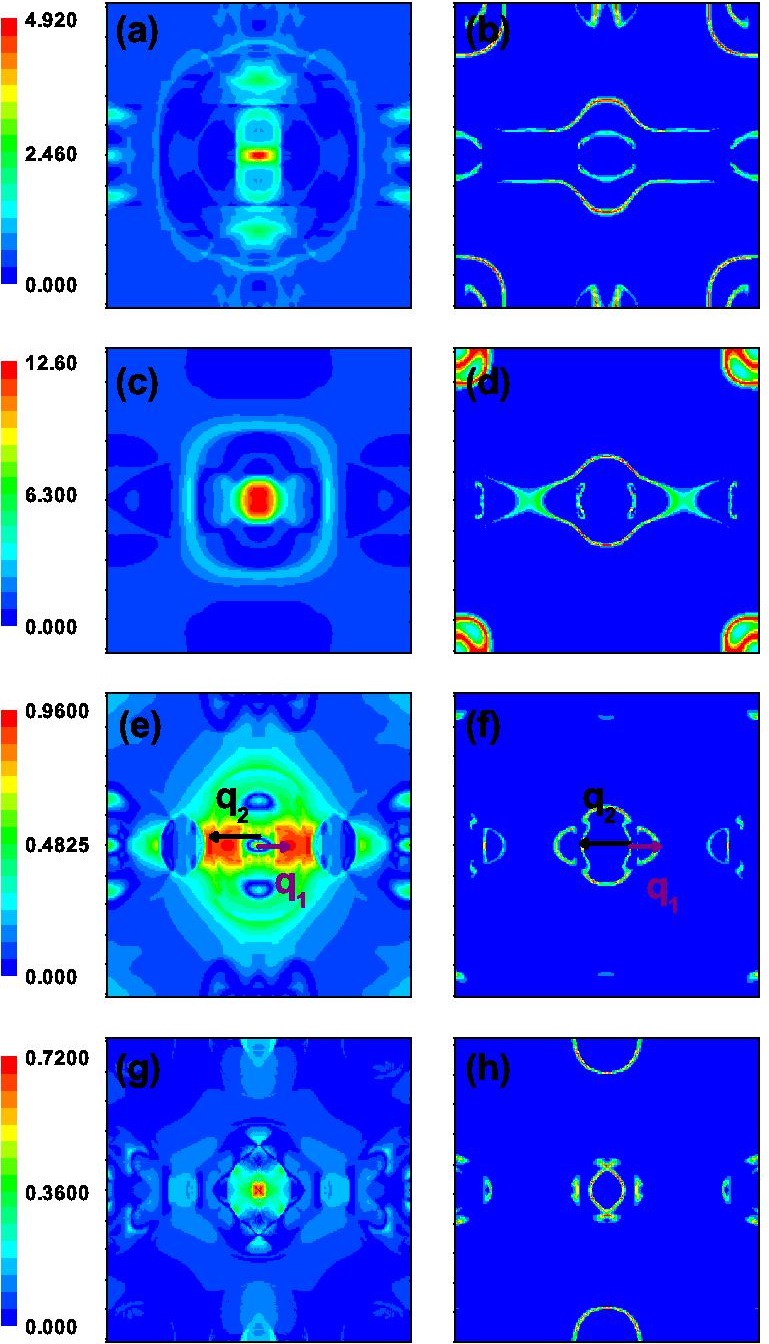}
\caption{(Color online) Energy evolution of the QPI
patterns in the SDW state. Panels (a), (c), (e) and (g) are
the QPI patterns at energies $\omega=-0.14$, $-0.08$, $-0.04$ and $0.08$,
respectively. The intensity maps of the corresponding quasiparticle
spectral functions are plotted in the right side panels.}
\label{f.2}
\end{figure}

Further analysis shows that the highly anisotropic FS induced by
the SDW is related directly to the orbital characters of the
normal state energy bands. As presented in Fig. \ref{f.1}(a), the
normal state FS is dominated by the $d_{xz}$, $d_{yz}$ and
$d_{xy}$ orbitals. The overlap between the $\Gamma$ and $X$
pockets are dominated by the $d_{yz}$ orbital, as a result, their
hybridization is maximum. Consequently, the coupling between the
inner $\Gamma$ pocket and the $X$ pocket gives rise to a large SDW
gap $\Delta_{y}\sim0.19eV$ along $k_{y}$ axis. On the other hand,
the SDW gap $\Delta_{x}\sim0.1eV$ along the $k_{x}$ axis is
smaller. This attributes to the fact that the overlap of the
$d_{xz}$ or $d_{xy}$ band between the $\Gamma$ pocket and the $X$
pocket is weak. However, we note that the SDW gap opens along the
full $k_x$ axis, while only at individual $k$ point along the
$k_y$ axis. In addition, the Dirac cone dispersions located at
$k_{x}$ axis are directly related to the orbital character of the
relative bands\cite{YRAN}. Thus, the orbital degree of freedom
plays an important role in the bands reconstruction in the SDW
state.
%Unlike the previous theories where the bands anisotropy
%relies on the ellipticity of the electron pockets\cite{KNOLLE},
%here it is related to the orbital characters of the relative bands.

%In the presence of impurity, the electrons are scattered.The
%ingoing electrons and outgoing electrons with the same energy will
%interfere with each other to give rise to standing waves in real
%space which can be directly observed by STM. Theoretically, the
%features of the QPI patterns can be qualitatively understood by
%the analysis of the joint density of states which is related to
%the constant energy contours(CECs).

Now, we study the energy evolution of the QPI patterns in the SDW
state and the results are presented in Fig.~\ref{f.2}. In
Fig.~\ref{f.2}(a), we show the QPI pattern for $\omega=-0.14$.
%It is clearly that the main intensity of this QPI pattern
%distributes within the momentum window denoted by the rectangle in
%panel (a).
Its main feature is obviously anisotropic with only $C_{2}$
symmetry and exhibits a dimer-like structure orienting along the
$q_{y}$ direction. We also show the intensity map of the
corresponding quasiparticle spectral functions(QSFs) at the same
energy in Fig.~\ref{f.2}(b). One can see that the electronic
states along the $k_{x}$ axis are fully gapped, which is the
consequence of the opening of the SDW gap $\Delta_x$ as can be
seen from the single-particle spectrum presented in Fig.
\ref{f.1}(d). We notice that the SDW gap does not exist along the
whole $k_{y}$ axis, although it opens at some $k_y$ points[Fig.
\ref{f.1}(d)]. Thus, the QSFs exhibit clear weights along the
$k_{y}$ axis, and the QPI pattern is dominated by the scattering
processes along the $k_{y}$ direction. It should
be noted that at the center region around the $\Gamma$ point the QPI
pattern exhibits like a short dimer. However, it is not the
substantial feature of the scattering pattern. When the energy is
slightly away from $\omega=-0.14$, the short dimer diminishes but
the vertical dimer structure remains. Actually, this dimer-like
structure of the QPI pattern persists in the energy window of
$-0.2<\omega<-0.12$, which coincides basically with the energy
region of the SDW gap in the single-particle spectrum as shown in
Fig. \ref{f.1}(d). With the increase of energy, the pattern
develops gradually into a broad peak around the $\Gamma$ point and
it has no obvious anisotropy as shown in Fig.~\ref{f.2}(c) for
$\omega=-0.08$. Moreover, the intensity of the QPI pattern is
nearly three times larger than that with $\omega=-0.14$. This is
the specific case in that this energy happens to be near the top
of the hole band around $(\pi,\pi)$ (not shown here), where the
density of states is large and consequently the QSFs show a
noticeable intensity around $(\pi,\pi)$ as shown in
Fig.~\ref{f.2}(d). Above this energy region, one approaches
gradually to the Fermi energy, so the QSFs will copy the main
features of the FS. In this case, the QPI pattern is dominated by
two sets of vertical dimers along the $q_y$ direction as indicated
by the $q_{1}$ and $q_{2}$ arrows in Fig.~\ref{f.2}(e) for
$\omega=-0.04$, which arises from the inter-hole-pocket and
intra-hole-pocket scatterings, respectively [see
Fig.~\ref{f.2}(f)]. What we distinguish the two
sets of vertical dimers comes from the features exhibited in the
QSF shown in Fig.~\ref{f.2}(f). From this figure, one can see that
the contour of the QSF is elongated along the $q_y$ direction, so
that the portions of the contours connected by $q_2$ become more or
less flat, and thus the scatterings across the vertical nearly
flat portions are dominant and give rise to the vertical dimers.
Actually, this kind of dimer structure of the QPI patterns
persists in the energy regime of $-0.03<\omega<-0.01$. In the
positive energy regime, the dimer structure of the QPI pattern
diminishes gradually. The typical QPI pattern is shown in
Fig.~\ref{f.2}(g) for $\omega=0.08$, which is dominated by a peak
around the $\Gamma$ point. This is consistent with the QSFs
analysis, where its anisotropy weakens significantly in the
positive energy regime [Fig.~\ref{f.2}(h)]. It is mainly because
the energy band above the Fermi level is less distorted by the SDW
formation, as can be seen from a comparison between
Fig.~\ref{f.1}(b) and (d). It was carefully checked that the above
mentioned features of the QPI patterns persist for weak impurity
scattering regardless of the sign of $V$.

Theoretically, the features of the QPI patterns in the SDW state have been investigated by previous studies\cite{QPIPRL,QPIPRB,QPIMAZIN}.
Similar to the results presented in Fig.~\ref{f.2}(e) and (f), the highly anisotropic QPI patterns obtained by previous studies\cite{QPIPRL,QPIPRB,QPIMAZIN}
are directly related to the topology of the SDW distorted constant
energy contours. Here, we emphasize another cause leading to the dimer structure in QPI pattern, that is the highly anisotropic SDW gaps
in the momentum space which is directly related to the orbital characters of the folded bands as addressed above.

\begin{figure}
\centering\includegraphics[width=3.2in]{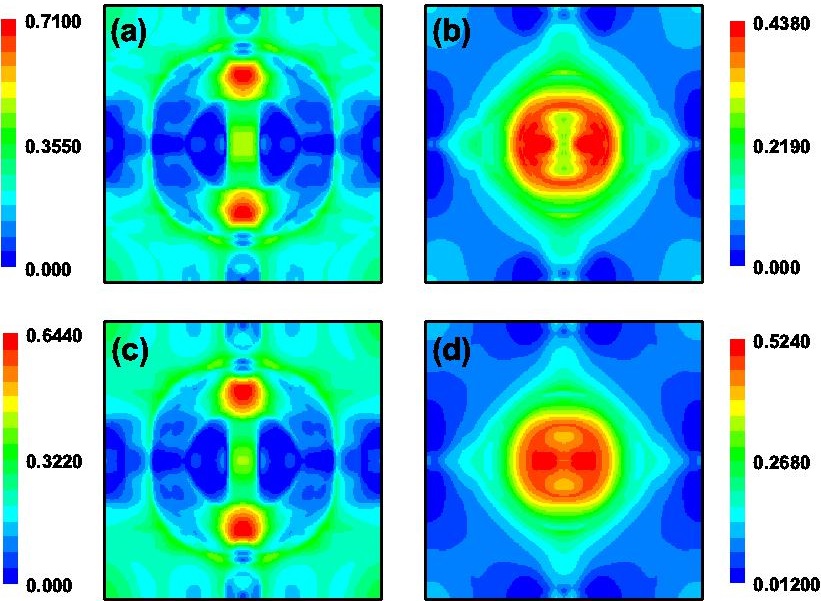}
\caption{(Color online) QPI patterns in the magnetically
driven nematic phase. Panels (a) and (c) are the QPI patterns at
$\omega=-0.14$ for $\xi=80$ and $\xi=20$, respectively. Panels
(b), (d) are the QPI patterns at $\omega=-0.04$ for $\xi=80$ and
$\xi=20$, respectively.}
\label{f.3}
\end{figure}

In the magnetically driven nematic phase, the long-range SDW order
gives way to the short-range antiferromagnetic correlation which
is modelled by the Lee-Rice-Anderson formula\cite{LEERICE} as
introduced above. In this case, the magnetic correlation length
$\xi$ is the relevant parameter. Therefore, let us study the
evolution of the QPI patterns with the reduction of the magnetic
correlation length $\xi$. The QPI patterns for $\omega=-0.14$ are
shown in Fig.~\ref{f.3}(a) and (c) for $\xi=80$ and $\xi=20$,
respectively. Compared to that with the SDW long-range order
[Fig.~\ref{f.2}(a)], the main feature still exhibits the
anisotropy with the noticeable intensity of the pattern along the
$q_y$ direction. It is the consequence of the gap opening due to
the short-range magnetic correlation similar to that in the SDW
scenario discussed above. The variation is that two new peaks
emerge at the two ends of the dimer-like structure. From
Fig.~\ref{f.3}(a) and (c), one can also see that this kind of QPI
structure is less affected by the change of $\xi$ from $80$ to
$20$. We also notice that it preserves in the same energy window
of $-0.2<\omega<-0.12$ to the case of the SDW order. It is interesting
to notice that similar QPI patterns have been experimentally observed in the
ferropnictide $122$ compounds\cite{CHUANG}, although the observed dimers are shorter. The results
for $\omega=-0.04$ are presented in Fig.~\ref{f.3}(b) and (d).
It shows that the $q_{1}$ and $q_{2}$ dimers presented in the SDW state
merge together to give rise to a loop around the $\Gamma$ point
when the correlation length is $\xi=80$. The mergence becomes
strongly with the further decreasing of $\xi$, and eventually a
nearly flat-top broad peak around the $\Gamma$ point will appear,
as has already been seen from Fig.~\ref{f.3}(d) for $\xi=20$.
Thus, it can be found that the anisotropy of the QPI patterns weakens significantly with the reduction of $\xi$. With
a finite correlation length $\xi$, the SDW gives way to the damped
spin excitations. The damping becomes stronger with the
decrease of the correlation length $\xi$. As a result, the
quasiparticle peak is broadened correspondingly due to its
coupling to these damped spin excitations as indicated by the
Lee-Rice-Anderson formula\cite{LEERICE} used above, so do the main
features of the QPI around the $\Gamma$ point.
In the case of
$\omega=-0.14$, the dimer structure along the $q_y$ direction in
the SDW phase is mainly caused by the opening of the SDW gap along
the $k_x$ direction of the single-particle spectrum which is
affected less by the quasiparticle damping. Thus, the anisotropic
feature of the QPI pattern for $\omega=-0.14$ persists even when
$\xi$ is significantly reduced.

\begin{figure}
\centering\includegraphics[width=3.2in]{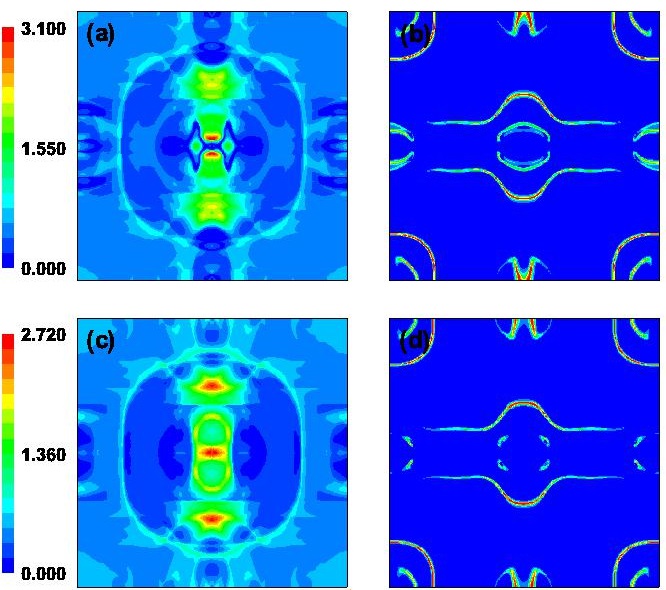}
\caption{(Color online) QPI patterns and intensity maps of QSFs in the SDW state with and without the
off-diagonal SDW orders. Panels (a) and (c) are the QPI patterns for
$\omega=-0.15$ without and with the off-diagonal SDW orders, respectively. The intensity maps of the corresponding QSFs are plotted in the right panels.}
\label{f.8}
\end{figure}

It was checked that the above mentioned features of the QPI patterns
are qualitatively unchanged when the interacting parameters vary within a realistic region suggested in Ref.\cite{FSLUO}. We have also considered the effects of the band structure on QPI patterns by using the LDA energy bands of LiFeAs\cite{LFAA,LFAB,LFAC}. It was
found that the QPI patterns also exhibit the prominent dimer structure when energy is in the SDW gap region. The dimers of the QPI patterns persist when the energy is around the Fermi level, and they weakens gradually with the reduction of the magnetic correlation length in the magnetic nematic phase. It should be noticed that there exist significant differences between the energy bands of LiFeAs and those of LaOFeAs\cite{GRASER,LFAA,LFAB}. Thus, the QPI features established above is robust.

Furthermore, it was checked that the main features of the QPI patterns remain when the off-diagonal SDW orders are considered.
Here, we do not present detail comparisons between the QPI patterns with and without the off-diagonal SDW orders in all cases. Instead, we give a typical example to show the behaviors of the scattering patterns when the off-diagonal SDW orders are taken into account. We show the QPI patterns for $\omega=-0.15$ in Fig.~\ref{f.8}(a) and (c) in the SDW state without and with the off-diagonal SDW orders, respectively. As shown in Fig.~\ref{f.8}(a), the QPI pattern for $\omega=-0.15$ exhibits a dimer-like structure orienting along the $q_{y}$ axis which is similar to that presented in Fig.~\ref{f.8}(c) with the off-diagonal SDW orders, although some differences around the $\Gamma$ point exists. This stems from the effect of the highly anisotropic SDW gaps in momentum space, which is less subjected to the inclusion of the off-diagonal SDW orders as shown by the QSFs in Fig.~\ref{f.8}(b) and (d). We also note that there is an approximate $10\%$ increase in magnitudes of the converged diagonal SDW orders after including the off-diagonal terms as presented in Sec.IIA. As a consequence, the QPI pattern shown in Fig.~\ref{f.8}(c) for $\omega=-0.15$ has nearly the same structure as that shown in Fig.~\ref{f.2}(a) for a slightly different energy of $\omega=-0.14$.

\subsection{QPI patterns in the orbital scenario of nematicity}
Experimentally, it was found that the degeneracy between the
$d_{xz}$ and $d_{yz}$ bands is lifted at a temperature coincident
with the onset of nematicity\cite{OODING,FESE,FEPG}. Theoretically, it was
proposed\cite{ONARI,OOCHEN,OOLV} that the nematicity may originate
from the orbital fluctuations. However, it is difficult to
distinguish the orbital driven nematicity from its spin
counterpart due to their mutual coupling\cite{NMREV}. In this
section, we study the QPI patterns induced by the potential
scattering in the orbital scenario of nematicity.

We note that it was recently revealed by ARPES experiments that
the orbital order is strongly momentum dependent in
FeSe\cite{FESED,FESEL}. With the help of the symmetry analysis, we
will first obtain the general form of the orbital order up to the
nearest-neighbor sites.
%We find that the orbital order induced QPI patterns exhibit a structural transition with the increase of energy, which can be
%used to identify the orbital driven nematicity in IBSCs.
In the present model, the point group of
the Fe plane is
$\{e,c_{4}^{1},c_{4}^{2},c_{4}^{3},c_{x},c_{y},\sigma_{+},\sigma_{-}\}$,
where $c_{4}^{n}=(c_{4}^{1})^{n}$ with $c_{4}^{1}$ a $\pi/2$ rotation along the $z$ axis followed by a Fe plane mirror reflection,
$c_{x}$ and $c_{y}$ are $\pi$ rotations along the $x$ and $y$ axis
of the Fe-Fe bond, respectively. $\sigma_{+}$ and $\sigma_{-}$ are
two mirrors determined by the nearest-neighbor Fe and As atoms,
respectively. It was found experimentally that the Dirac cone dispersion exists in the SDW state\cite{EDIRAC}.
As pointed out by previous study\cite{YRAN}, the emergence of the Dirac cones is a consequence of the $c_{x}$ and $c_{y}$ symmetries.
In this way, we expect that the $c_{x}$ and $c_{y}$ symmetries are preserved in the SDW state. As proposed theoretically\cite{CKXU,CFANG,ONARI,FERNANDES},
the SDW state is expected to occur as a result of the spontaneous symmetry breaking of the nematic phase.
Thus, the $c_{x}$ and $c_{y}$ symmetries are preserved in the nematic phase. This is further supported by
recent ARPES observations\cite{DLFENG,ZXSHEN}. In this way, the
point group of the nematic phase is $\{e,c_{4}^{2},c_{x},c_{y}\}$.
Generally, the Hamiltonian for the orbital order can be written as
$H_{orb}=\sum_{ij,ab,\sigma}\lambda_{ij,ab}C^{+}_{ia\sigma}C_{jb\sigma}$,
where $i,j$ and $a,b$ are the lattice sites and orbital indices,
respectively.
%The orbital order is of a $d$-wave form under the $c_{4}^{1}$ rotation, meanwhile it shall obey the $c_{x}$, $c_{y}$
%and $c_{2}$ symmetries. Furthermore, we restrict ourself to the
Considering the point group symmetry and that the $t_{2g}$
orbitals dominate the low energy electronic bands, we find
that the $H_{orb}$ with only the on-site orbital order reads
$\sum_{i,\sigma}\lambda_{0}(C^{+}_{i\sigma,xz}C_{i\sigma,xz}-C^{+}_{i\sigma,yz}C_{i\sigma,yz})$.
Other terms, such as
$\sum_{i\sigma}C^{+}_{i\sigma,xz}C_{i\sigma,yz}$, are forbidden
since they break the $c_{x}$ and $c_{y}$ symmetries. The $H_{orb}$
with the orbital orders up to the nearest-neighbor sites are
written as $\sum_{k\sigma}\lambda_{0}(C^{+}_{k\sigma,xz}C_{k\sigma,xz}-C^{+}_{k\sigma,yz}C_{k\sigma,yz})+\lambda_{1}(\cos k_{x}+\cos
k_{y})(C^{+}_{k\sigma,xz}C_{k\sigma,xz}-C^{+}_{k\sigma,yz}C_{k\sigma,yz})+\lambda_{2}(\cos
k_{x}-\cos
k_{y})(C^{+}_{k\sigma,xz}C_{k\sigma,xz}+C^{+}_{k\sigma,yz}C_{k\sigma,yz})+i\lambda_{3}\sin
k_{x}C^{+}_{k\sigma,xy}C_{k\sigma,xz}-i\lambda_{3}\sin
k_{y}C^{+}_{k\sigma,xy}C_{k\sigma,yz}+h.c.+\lambda_{4}(\cos
k_{x}-\cos k_{y})C^{+}_{k\sigma,xy}C_{k\sigma,xy}$. We can of
course construct the orbital orders up to the next-nearest-neighbor
sites in a similar way, however, this introduces more parameters
which complicate the problem. In general, the magnitude of the
order will decrease with the increase of the lattice distance. So,
we will focus on the orbital orders up to the nearest-neighbor
sites which are believed to have captured the main physics.

\subsubsection{orbital order with momentum dependence}
\begin{figure}
\centering\includegraphics[width=3.0in]{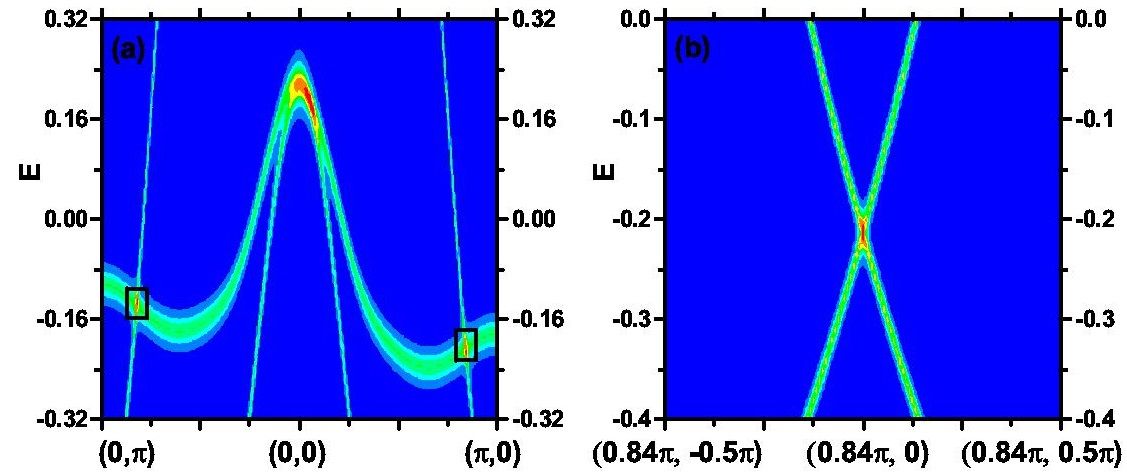}
\caption{(Color online) (a) Energy bands along the high symmetry
directions in the presence of momentum dependent orbital order.
The outer hole band along the $(0,0)-(\pi,0)$ and $(0,0)-(0,\pi)$
directions are mainly composed of the $d_{yz}$ and $d_{xz}$ orbitals, respectively.
The two black squares mark the location of the two Dirac nodes
which reside on the $k_{x}$ and $k_{y}$ axes. (b)
Linear dispersion for the Dirac cone located at $k_{x}$ axis. This
panel is plotted along the $(0.84\pi,k_{y})$ line. The Dirac node
locates at $(0.84\pi,0)$.} \label{koob}
\end{figure}

%Recently, it was found the splitting energy between the $d_{xz}$ and $d_{yz}$ bands is strongly momentum dependent\cite{FESED,FESEL}.
We use the above obtained Hamiltonian for the orbital orders to
fit the splitting energy between the $d_{xz}$ and $d_{yz}$ bands
observed in experiments\cite{FESED,FESEL}. Experimentally, the splitting at $\Gamma$ point is about $0.02$, and that between the $d_{yz}$ band at $X$ point and the $d_{xz}$ band at $Y$ point is increased to be $0.08$. The parameters are
$\lambda_{0}=0.01,\lambda_{1}=0,\lambda_{2}=0.015,\lambda_{3}=0.005,\lambda_{4}=0.002$,
which can reproduce the experimental data well. The resulted energy bands are shown in Fig.~\ref{koob}(a).
The outer hole band along the $(0,0)-(\pi,0)$ direction is mainly of the $d_{yz}$ orbital character, while that  along the $(0,0)-(0,\pi)$ direction is of $d_{xz}$ character.

Let us first discuss the effects on the energy band due to the
orbital orders. Because the outer hole band and the electron band
along the $k_x$ axis are even and odd parities under the $c_{x}$ operation. As a result, no gap opens at the intersection
point $(0.84\pi,0)$ between the outer hole band and the electron band. It leads to a Dirac cone
dispersion around $(\pi,0)$. Equivalently, there exists another
Dirac cone band around $(0,\pi)$. The locations of the two Dirac
points are marked by the black squares in Fig.~\ref{koob}(a).
These two Dirac cones are inequivalent in energy due to the
$C_{4}$ symmetry breaking induced by the orbital orders. The node
energies are $\omega_{x}=-0.21$ and $\omega_{y}=-0.14$ for the
Dirac cones located at the $k_{x}$ and $k_{y}$ axes, respectively.
In Fig.~\ref{koob}(b), we show the dispersion of the low energy
Dirac cone at $(0.84\pi,0)$ along the $(0.84\pi,k_{y})$ direction. It is interesting to notice that the similar cone like dispersion has
been observed by the recent ARPES measurements on
thin FeSe films\cite{DLFENG,ZXSHEN}, although the observed Dirac cone is
just below the Fermi level. On the other hand, compared to the
complicated energy bands reconstruction induced by the SDW order
as shown in Fig.~\ref{f.1}(c), the distorted energy bands due to the
orbital orders are dominated by two hole bands around the $\Gamma$ point and
two electron bands around the $(\pi,0)$ and $(0,\pi)$ points.

%With the orbital order shown above, we study the energy evolution of the QPI patterns.
%The scattering patterns are obtained in a wide energy window from $\omega=-0.3$ to $\omega=0.3$,
%and their anisotropy are rather weak for $\omega>-0.06$ which is significantly above the node energies.
%This indicates that the anisotropic features of the QPI patterns are associated with the Dirac core dispersions.

\begin{figure}
\centering\includegraphics[width=3.2in]{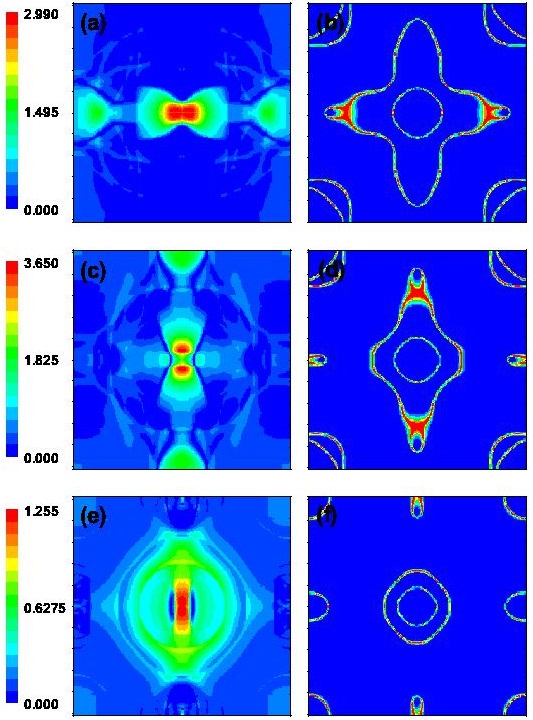}
\caption{(Color online) Energy evolution of the QPI patterns in
the orbital driven nematic phase with strongly momentum dependent
orbital order. Left panels are the QPI patterns for
$\omega=-0.24$, $-0.18$ and $-0.1$, respectively. Right panels are
the intensity map of the corresponding quasiparticle spectral
functions.}
\label{kqpi}
\end{figure}

Now, we turn to the study of the energy evolution of the QPI
patterns in the orbital scenario of nematicity. The results are
obtained in a wide energy window from $\omega=-0.3$ to
$\omega=0.3$, and the anisotropy of the QPI pattern is rather weak
when $\omega>-0.06$ suggesting that it is irrelevant to the
orbital orders above this energy. Thus, we show the typical
patterns in Fig.~\ref{kqpi} (a), (c), and (e) for $\omega=-0.24$, $-0.18$ and $-0.1$,
respectively. For $\omega=-0.24$, the scattering pattern is
highly anisotropic with its main intensity along the $q_{x}$ axis, leading to a horizontal dimer.
When the energy is increased to
be near $\omega=-0.18$, the QPI pattern turns to be a vertical
dimer. This vertical dimer remains up to a higher energy
$\omega=-0.1$ which is slightly above $\omega_{y}$, although
additional structures develop around the dimer. With the further
increasing of energy, the anisotropy of the QPI patterns weakens
significantly. Especially, the scattering patterns exhibit little
anisotropy in the positive energy region.

%This indicates that the anisotropic features of the QPI patterns
%are associated with the Dirac core dispersions.

The transition of the QPI patterns from the horizontal to vertical
structure is associated with the inequivalent energy positions of
the two Dirac cones due to the $C_{4}$ symmetry breaking induced
by the orbital orders. As shown in Fig.~\ref{kqpi}(b), (d) and (f),
the spectral function shows a large intensity around the Dirac cones. It attributes to the small velocity of the
outer $d_{yz}$ and $d_{xz}$ bands around the two Dirac cones.
So, the QPI process is dominated by the Dirac cone to Dirac cone
scatterings of quasiparticles. For $\omega=-0.24$, the nearby Dirac cones
situate at the $k_x$ axis. But, for $\omega=-0.18$ and -0.1, they
situate at the $k_y$ axis. So, the Dirac cone to Dirac cone
scatterings lead to the transition of the QPI patterns which is
related to the presence of the orbital orders.

\subsubsection{orbital order without momentum dependence}
\begin{figure}
\centering\includegraphics[width=3.2in]{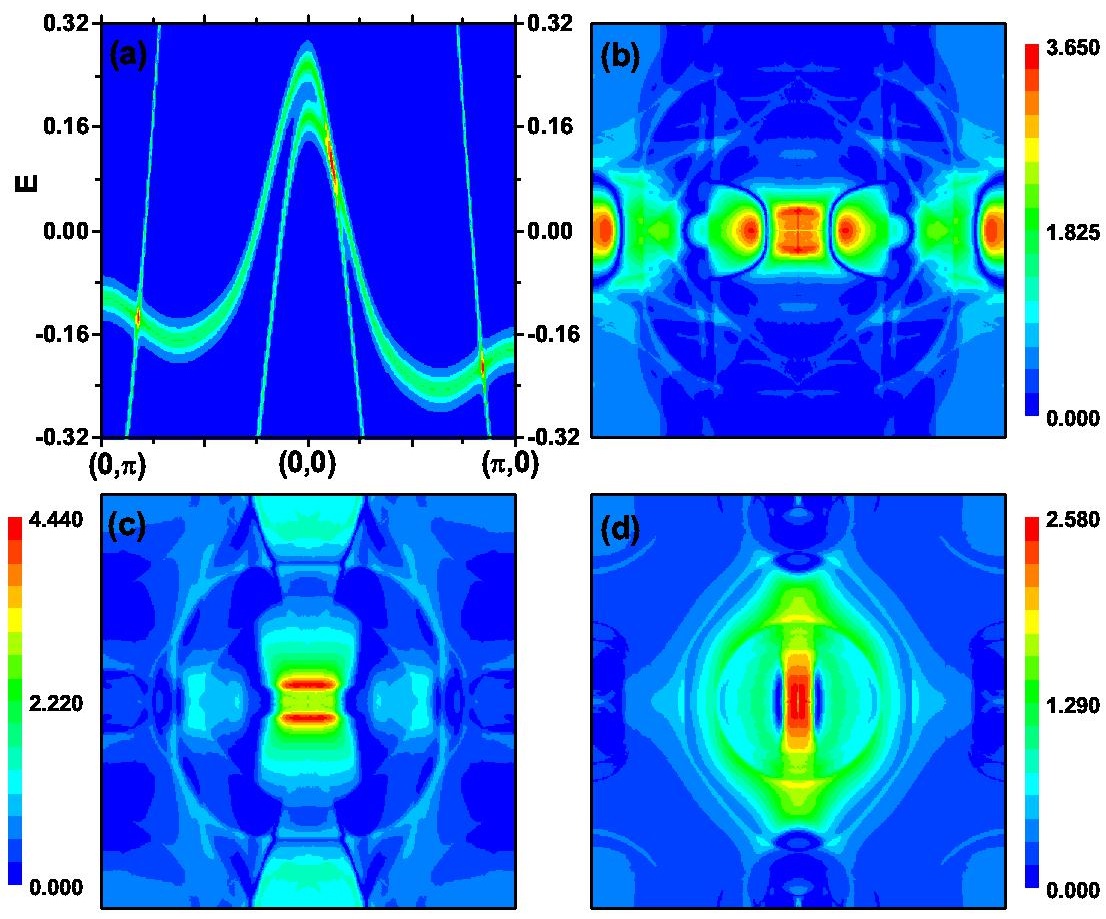}
\caption{(Color online) Energy bands reconstruction and the energy evolution of the QPI patterns in
the orbital driven nematic phase with the momentum independent
orbital order. Panel (a) is the energy bands along the high symmetry direction. The
outer hole bands along the $(0,0)-(\pi,0)$ and $(0,0)-(0,\pi)$ directions are mainly of the $d_{yz}$ and $d_{xz}$ characters,
respectively. Panels (b), (c) and
(d) are the QPI patterns for $\omega=-0.24$, $-0.18$ and $-0.1$, respectively.}
\label{f.5}
\end{figure}

Though the orbital order is momentum dependent in
FeSe\cite{FESED,FESEL}, this may vary in different kinds of IBSCs.
As a comparison, we will study the energy evolution of the QPI
patterns with a momentum independent orbital order in this
section. The parameters
$\lambda_{0}=0.04,\lambda_{1,2,3,4}=0$
give rise to a momentum independent orbital order as that used in
previous studies\cite{OOCHEN,OOLV}. As shown in Fig.~\ref{f.5}(a),
the resulted energy splitting between the $d_{xz}$ and $d_{yz}$
bands are $0.08$ at $\Gamma$. And it has the same splitting energy between the $d_{xz}$ band around $Y$ and
the $d_{yz}$ band around $X$. The
magnitude of this splitting energy is comparable to that obtained
by the ARPES experiments\cite{FEPG,APZHANG}.

%\begin{figure}[tbp]
%\vspace{-0.01in}
%\centering\includegraphics[width=3.5in]{FEOOBAND.EPS}
%\caption{(Color online) Energy bands along the high symmetry
%direction in the scenario of the momentum independent orbital
%order.} \label{f.4}
%\end{figure}

To make a comparison to the results for the momentum dependent
orbital order, we present the QPI patterns at $\omega=-0.24$,
$-0.18$, and $-0.1$ in Fig.~\ref{f.5}(b)-(d). Quite similarly, the main feature
of the QPI patterns exhibits a transition from the horizontal
dimer at a low energy $\omega=-0.24$ to the vertical dimer at
$\omega=-0.18$, and the vertical dimer persists up to a high
energy $\omega=-0.1$. It should be noticed that the horizontal dimers with the highest intensity
in Fig.~\ref{f.5}(c) are not the substantial feature of the QPI pattern due to the fact that
they diminish when the energy is away from $\omega=-0.18$.
It is carefully checked that the main
features of the QPI patterns keep unchanged when the orbital
orders evolve smoothly from the strongly momentum dependent case
to the momentum independent case, if the splitting energy between the $d_{xz}$ band
at $Y$ and the $d_{yz}$ band at $X$ is fixed to be $0.08$.
Thus, the above established transition of the
QPI patterns is a robust phenomenon, regardless of the momentum
dependence of the orbital orders. The difference in the energy
splitting of the $d_{xz}$ and $d_{yz}$ bands between the momentum
dependent and independent orbital orders is that there is a constant
energy splitting in the latter case.
Thus, the two Dirac cones that situate near the $(\pi,0)$ and
$(0,\pi)$ points are not affected by the difference of the
momentum dependence of the orbital order
%as long as the splitting energy is fixed between the $d_{xz}$ band at $(0,\pi)$ and the $d_{yz}$ band at $(\pi,0)$,
as
shown clearly in Fig.~\ref{koob}(a) and Fig.~\ref{f.5}(a). Thus, the main feature of the QPI
patterns is unchanged because it mainly arises from the Dirac cones
to Dirac cones scatterings.

Before concluding Sec.IIIB, it is worth mentioning that the transition of the QPI patterns is
related to the $C_{4}$ symmetry breaking induced by the orbital
order and the parities of relative bands which are both involved
with the symmetry of the orbital nematic phase. Thus it is expected that the above established
transition remains unchanged when the interactions between
electrons are taken into account. Actually, there exists numerical
evidence\cite{FERBER} that the symmetries of relative bands do not
change when the correlation effects are considered. We also note that the established transition of QPI patterns occurs at energies much below the Fermi level.
Considering that the bandwidth is significantly reduced when
the Coulomb interactions are taken into
account\cite{FERBER,AICHHORN,YIN,WERNER}, we expect that the energy difference between the transition
and the Fermi level will be reduced and it will facilitate the experimental measurements.
On the other hand, it was experimentally found on the FeSe films that
the Dirac cones reside slightly below the Fermi level\cite{DLFENG,ZXSHEN}. This indicates
that the transition of QPI patterns can be observed at energies close to
the Fermi level in some iron pnictides, at least in the FeSe films.

\section{summary and discussion}
In conclusion, we have studied theoretically the quasiparticle
scattering interference patterns in the nematic phase of the
iron-based superconductors based on the magnetic and orbital
scenarios, respectively.

Deep in the SDW state,
the QPI patterns exhibit a dimer structure in a wide energy region and develop a bi-dimer structure when the
energy is increased to be near the Fermi level. It is also shown
that the dimer structure of the QPI patterns still exists when the SDW state is replaced by the state with
fluctuating magnetic order, though the bi-dimer structure is
smeared due to the mergence of the two sets of dimers when the short-range correlation length is significantly decreased.

Thus, we identify that the QPI dimers present in the SDW and the magnetic nematic phase orient along the ferromagnetic direction of the SDW order as long as the correlation length is not significantly decreased. Our results based on the magnetic scenario of nematicity are qualitatively consistent with the STM observations on NaFeAs\cite{QPIN}.

In the orbital scenario of nematicity, the QPI patterns are dominated by a dimer structure along the $q_x$ or $q_y$ axes in a wide energy region. A $\pi/2$ rotation of the dimer structure occurs when the energy
increases from the lower Dirac node to the higher one. Above the two Dirac nodes, the anisotropy of the QPI patterns is significantly weakened. Furthermore, it is found that this
transition is insensitive to the momentum dependence of the orbital order.

Theoretically, the suggestions for distinguishing the SDW order from the orbital driven nematicity by using STM technique have been proposed\cite{MONM,JPSJ}. Plonka {\it et al.}\cite{MONM} focus on the contributions of the band extrema (local maxima or minima in the bands) to the local density of states (LDOS) and QPI. They found that the features related to the differences between two band extrema which track the orbital splitting or the SDW gap can be used to distinguish between the orbital splitting and the SDW order. The features are easy to be detected in LDOS, but are obscured in QPI by contributions from orbitals other than $d_{xz}$ and $d_{yz}$. Eremin {\it et al.}\cite{JPSJ} focus on the QPI at an energy near the Fermi level, so the anisotropic QPI patterns are closely related to the topology of the SDW or orbital order distorted constant energy contours near the Fermi level. Here, we uncover the significant contributions to QPI from the Dirac cone to Dirac cone scatterings of quasiparticles in the orbital scenario. For the magnetic scenario, we elaborate the robust dimer structure in QPI patterns resulting from the highly anisotropic SDW gaps in the momentum space which are directly related to the orbital characters of the folded bands.
We further show that due to the inequivalent energies of the two Dirac nodes, the dimer structure in the orbital scenario undergoes a $\pi/2$ rotation with the increase of energy, which contrasts clearly with the case of the magnetic scenario.
These features of the QPI patterns have not been addressed in previous studies\cite{QPIPRL,QPIPRB,QPIMAZIN,JPSJ,MONM}. In addition, compared to Ref.~\cite{JPSJ} in which the two-orbital model is used, we carry out the calculations by using the more realistic five-orbital model.

%The obtained intrinsic differences of the scattering patterns between the spin and the
%orbital cases are sufficient to be used to distinguish the origin of nematicity in iron pnictides.}

From the results presented in this paper, we conclude that the QPI patterns exhibit dimer structure in a wide energy region in both the magnetic and the orbital scenarios of nematicity. The dimer tends to orient along the ferromagnetic direction in the magnetic case. However, it undergoes a $\pi/2$ rotation with the increase of energy in the orbital case. Thus, our results established in this paper for the QPI
patterns may be used to probe the origin of nematicity in various
iron-based superconductors.

%Generally, the magnetic and orbital fluctuations couple with each
%other in the nematic phase. Our study in this paper apply to the
%scenario where the driven force of namaticity is either magnetic
%or orbital fluctuations dominated. In this way, either magnetic or
%orbital fluctuations dominates the main features of the QPI
%patterns. As emphasized by previous study\cite{NMREV}, the
%question of the origin of nematicity is meaningless if the orbital
%and magnetic fluctuations strongly coupled with each other.

%In contrast, the experimental observations\cite{QPIN,QPIP} are in
%qualitatively consistent with the results obtained within the
%magnetic scenario of nematicity. This indicates that the
%nematicity are magnetic driven in NaFeAs. Our results can be used
%as a probe to identify the origin of namaticity in various IBSCs,
%especially the widely studied FeSe\cite{FESE,DLFENG,ZXSHEN}.

\section{acknowledgement}
This work was supported by the National Natural
Science Foundation of China (11190023 and 11374138), and National Key Projects for Research $\&$ Development
of China (Grant No. 2016YFA0300401).

\end{document}